\documentclass[11pt, dvips]{article}
\def\gsim{\raise0.3ex\hbox{$>$\kern-0.75em\raise-1.1ex\hbox{$\sim$}}}
\begin{document}
\begin{center}
{\LARGE \bf Weak doubly special relativity \\
and ultra-high energy cosmic ray experiments}
\end{center}
\vskip 2mm
\begin{center}
%
%
{\bf Luis Gonzalez-Mestres\\}
{\it L.A.P.P., B.P. 110, 74941 Annecy-le-Vieux Cedex, France}
\end{center}
\vskip 2mm
\begin{center}
{\large \bf Abstract\\}
\end{center}
\vspace{-0.5ex}
{\bf Should projects of space experiments on ultra-high energy cosmic rays be supported, whatever AUGER results will turn out to be? We claim that this is indeed the case} and that it is a real need, from a scientific point of view, to perform these cosmic-ray studies with an expected effective detection area solid angle efficiency product above 2 x $10^5 Km^2 sr$ . \\
It is now widely admitted that models of Lorentz symmetry violation (LSV) at the Planck scale based on power-like extrapolations down to cosmic-ray scales and able to account for a possible absence of the Greisen-Zatsepin-Kuzmin (GZK) cutoff exist and require the existence of a privileged inertial rest frame, as we proposed in 1997 (paper physics/9704017 and subsequent work). The favoured energy dependence of the LSV parameter will then be quadratic rather than linear, at least as seen from the energy region below $10^{20}$ eV. This approach (weak doubly special relativity, WDSR) is different from the version of doubly special relativity defended by several authors, where the laws of Physics are required to be exactly identical in all inertial reference frames (strong doubly special relativity, SDSR). The Kirzhnits-Chechin model (1971) is a particular case of SDSR and, as first pointed out in our hep-th/0210141 paper, it does not lead to a suppression of the GZK cutoff.\\
Another attempt to explain a possible absence of the GZK cutoff was the model of hadronic matter by Saso and Tati (1972). Like WDSR, it postulates the existence of a privileged rest frame. But, contrary to WDSR, it does not assume any deformation of kinematics and its basic hypothesis is
 a property of vacuum suppressing hadrons above a certain Lorentz factor. Such a strong conjecture, which is not based on any precise dynamical model, seems unlikely and is not consistent with WDSR or quantum gravity patterns where in all cases the possible absence of the GZK cutoff is obtained by comparing the deformation energy with that of Cosmic Microwave Background (CMB) photons. \\
We therefore conclude that, to date, WDSR patterns based on a deformation of special relativity with a privileged (vacuum) rest frame are the only clear and consistent candidate to explain a possible absence of the GZK cutoff invoking deviations from standard relativity. Testing such scenarios requires a performant experimental study of the highest energy cosmic-ray air showers. \\
It is also to be emphasized, as in hep-ph/0510361 , that the usual hypothesis of a power-like dependence of the LSV effective parameters not being altered by any intermediate energy scale is not the only possible one. More sophisticate energy-dependences are also to be considered and may even be natural, involving significant effective thresholds at intermediate energies. Such thresholds may exist between the Planck scale and the highest cosmic-ray energies, or between ultra-high cosmic-ray (UHCR) energies and the TeV scale, and play an important role.
Predictions of LSV patterns can be dramatically modified and phenomena excluded at an energy scale space may manifest themselves at higher energies. Therefore, experiments sensitive to UHCR energies as high as possible become necessary irrespective of AUGER results. By combining both kinds of experiments, future results of cosmic-ray observations will hopefully be able to accurately test, for a large family of models involving various patterns of Planck-scale physics, the validity of special relativity at ultra-high energies as well as the possible existence of an absolute local rest frame in the real world.
\vskip 3mm
Contribution to the I Symposium on European Strategy for Particle Physics, Orsay (France), January 30 - February 1st , 2006 , supporting a statement in favour of UHECR space experiments.

\vspace{1ex}

~

~
\section{Introduction}

Contrary to naive expectations, the theoretical study of possible violations of special relativity at ultra-high energy (UHE) has shown that data on the highest energy cosmic rays can provide a unique and powerful microscope directly focused on Planck-scale physics. Tracks of physics at the Planck scale, even becoming as small as $\approx 10^{-20}$ as compared to the basic observables in the $10^{19}~ - ~ 10^{20} ~eV ~$ region, can still generate important effects and phenomena directly testable by UHECR experiments. It is a unique physics opportunity which, by itself, justifies the study of this energy region by large-scale experiments like AUGER \cite{AUG} as well as by future space experiments to get more precise data at the highest available cosmic-ray energies. In this respect, it must be kept in mind that space experiments can provide geometry factors exceeding by a factor as large as 30 that of any Earth-based experiment \cite{OWL}. Improving the measured UHECR flux by such a factor may be equivalent to winning almost an order of magnitude in available energies. This would certainly be a crucial progress in the search for tracks of Planck-scale physics in UHECR, not only because experimental data would get significantly closer to that scale but also because the events observed from a satellite will be seen from a different perspective as compared to Earth-based detectors.

By now, preliminary AUGER data \cite{AUG05} seem not to exclude a possible absence of the Greisen-Zatsepin-Kuzmin (GZK) cutoff \cite{GZK1,GZK2} . In this case, Lorentz symmetry violation at the Planck scale \cite{gon97a} would be a natural candidate to explain such an observation provided it is assumed that a privileged rest frame exists (the {\it vacuum rest frame}). But it may also happen that future data with better statistics show the existence of such a cutoff. This would not rule out all LSV patterns and tests as precise as possible of Lorentz symmetry above $10^{20} ~ eV$ would still be necessary.

The model we proposed in 1997 in \cite{gon97a} , and also in \cite{gon97b,gon97c} and in other papers of the same period (see {\it arXiv.org }), is indeed able to explain a possible absence of the GZK cutoff, contrary to that considered previously by Kirzhnits and Chechin \cite{Kir} which was shown \cite{gon02a,gon02b} not to be able to account for such an effect. The reason is that the Kirzhnits-Chechin model is a form of STRONG doubly special relativity (SDSR), where it is required that the laws of Physics be exactly the same in all inertial frames \cite{Amel04a,Smolin05a}. This approach assumes that the action is invariant under space-time diffeomorphisms, forbidding the appearance of a preferred reference frame. It turns out to imply a substantial violation of standard energy-momentum conservation rules in the physical inertial frames and precludes in this way the effect announced by the authors of \cite{Kir} . 

As stressed in \cite{gon05} , models like that proposed in \cite{gon97a} can fit into the larger family of modified special relativity patterns that we call WEAK doubly special relativity, WDSR. In WDSR, it is still required like in SDSR that the laws of physics be described by two universal constants, {\it c} and the fundamental length {\it a}. But the equations of motion are not identical in all inertial frames: they are identical in the $a ~ \rightarrow ~ 0 $ limit, are almost identical (identical up to the form of the deformation term) at energies well below Planck scale and tend to standard special relativity as the momentum scale decreases. The vacuum rest frame can then, in the examples we consider, be characterized by the isotropy of the laws of Physics. The universality of these laws remains in WDSR, up to the boost between the inertial frame considered and the vacuum rest frame. In the case of kinematics for free particles, such a boost is a linear energy-momentum transformation, preserving the additiveness of these four observables. Similar patterns are assumed for more involved laws of Physics. {\bf If the GZK cutoff does not exist in the real world and if deviations from standard relativity are the right explanation to its absence, our LSV patterns of the WDSR type \cite{gon05} can account for such an effect, but not those following the SDSR criteria.} Therefore, if the GZK cutoff is not there and the explanation is a deviation from standard relativity, this will be a proof of the existence of a privileged (vacuum) rest frame in our Universe.

An example of WDSR can be provided by a dispersion relation of the type (1) - (3) presented in {\bf Section 2} (and already used in our 1997 papers), assumed to be valid in the vacuum rest frame, and by its (linearly) Lorentz-transformed sets of equations in the other inertial frames. This is a classical illustration of deformed relativistic kinematics (DRK) in WDSR. 

{\bf In WDSR and contrary to SDSR, the measured energy and momentum can be additive for sets of free particles and conserved. This has always been 
our working hypothesis, contrary to SDSR calculations. Although 
DRK is a direct consequence of the onset of an effective fundamental length scale, $a$ , the WDSR space-time can remain continuous and translation invariance can still hold at all scales.} Even if our April 1997 paper \cite{gon97a} used a parameterisation inspired by the Bravais lattice, the discreteness of space was not an essential property and the concept was readily generalized to any model involving short-range correlations related to a fundamental length scale (see f.i. \cite{gon97b} ). 

Expressions like (1) - (3) involve a power-like extrapolation of the effective LSV coefficient, from the Planck scale down (at least) to all scales above the highest mass scale of the particles considered. As emphasized in \cite{gon05} , this is only the simplest possible scenario and other forms of energy-dependence are possible. It must also be noticed that the WDSR models proposed in our papers since 1997 are not field-theoretical in the sense considered by \cite{coll04} . They therefore escape the criticism formulated by these authors. Furthermore, because of the predicted collapse of the final-state phase space at ultra-high energy \cite{gon97f} , which would in particular be at the origin of the absence of GZK cutoff, implementation of unitarity at these energies can undergo important changes in WDSR as compared to standard scenarios. This important question will be discussed elsewhere. 

Also, the general requirement that the deformation term of WDSR be negative in order to
prevent spontaneous decays of ultra-high energy (UHE) particles in vacuum was first formulated and studied in our 1997
papers (f.i. \cite{gon97e}) where possible violations of the equivalence principle were explicitly discussed and the question of the universality of the coefficient of the deformation term was equally dealt with \cite{gon97b,gon97c,gon97d, gon97g}. It was made clear there that this universality is not possible for large bodies, for which a different law is needed where the deformation coefficient depends on the mass of the object considered \cite{gon97d,gon99}. For elementary particles, the universality of the deformation appeared as a natural but not {\it a priori} compulsory hypothesis, often imposed in practice to a good approximation by phenomenological considerations. The analysis was further developed in \cite{gon00a} to obtain
bounds on standard LDRK
parameters (see {\bf Section 3}) and led in this case \cite{gon00a,gon00c} to 
$E_{QG} ~ \gsim ~ 10^{26} ~ GeV$ where $E_{QG}$ is the effective quantum gravity scale or any other relevant fundamental scale playing a similar role (the inverse of the effective fundamental length scale $a$).

In what follows, the word "deformation" stands for a set of special-relativity violating terms which tend to zero (possibly up to very small constants) in the infrared limit, faster than the conventional squared momentum term of relativistic kinematics. This may correspond to the existence of a preferred inertial frame as in WDSR, or to a pattern without a vacuum rest frame (SDSR). Usually, like in (1) - (3) , the deformation is power-like down to the particle mass scales and involves no intermediate energy thresholds. However, such thresholds may exist and play a significant role, as discussed in \cite{gon05} and in the present paper. Previous examples of deformations of relativistic kinematics not following a uniform power law were already studied in our articles on superluminal particles such as \cite{gonsl97} and had been proposed before in our 1995-96 papers. The pattern based on a non-universality of the critical speed in vacuum, used by Coleman and Glashow \cite{CG98} to reproduce our April 1997 results \cite{gon97a} on the absence of GZK cutoff and the stability of unstable particles at very high energy, is not of the DRK type as the non-universality of the critical speed remains unchanged at small momenta.

As it will be explicitly reminded later, the DRK patterns we consider do not contain any intrinsic inhibition of the existence of hadronic matter. A deformation of relativistic kinematics generated at the Planck scale and decreasing by a power law is enough to produce dramatic effects well below Planck scale, at energy scales where it has became very small numerically: around $\approx 10^{-18}$ , as compared with the standard momentum term, at $E ~ \simeq ~10^{20} ~ eV$ for the example considered in \cite{gon97a} . The only mechanism at work are kinematical balances where such extremely small terms appear to be enough to change or produce observable phenomena. Being so small and following a power law, they escape all possible low-energy tests of relativity. Our DRK models require only rather natural modifications of relativistic kinematics, contrary to the model of hadronic matter by Saso and Tati \cite{SATO}. Like WDSR, Sato and Tati postulated the existence of a privileged rest frame. But, contrary to WDSR, they did not assume any deformation of kinematics. Instead, their basic hypothesis was a strong dynamical one: a property of the present vacuum suppressing hadrons above a certain Lorentz factor. This conjecture was not based on a precise dynamical model and is not consistent with WDSR or quantum gravity patterns where in all cases the possible absence of the GZK cutoff is obtained by a simple kinematical balance, comparing mainly the deformation energy of the incoming cosmic ray with the energy of the Cosmic Microwave Background (CMB) photon. To suppress the GZK cutoff, WDSR does not require any dynamical restriction on the possibility for hadrons to exist or be formed in the physical vacuum and, contrary to the Sato-Tati model, it even makes possible the stability of the neutron above $10^{20}
 ~ eV$ . Actually, Sato and Tati formulated a strong constraint in order to forbid 
the formation of the most significant region of the possible final states of the interaction of a UHECR with a CMB photon and, as this constraint could not be made Lorentz invariant, they were led to require a violation of Lorentz symmetry. To date, nothing seems to support the basic dynamical hypothesis of the Sato-Tati model on hadronic matter. Furthermore, consistency with data is far from obvious, as: a) if the observed UHECR are protons, they seem to escape the Sato-Tati constraint; b) if there are nuclei, their Lorentz factors lie below the suppressed region; c) there are other hadronic processes than just signe pion formation.

~

~
\section{QDRK with a vacuum rest frame \\ and energy-momentum conservation}

The role of possible LSV in astrophysical processes at very
high energy has been discussed and updated in \cite{gon00a,gon00b} , and later in \cite{gon03a,gon03b} and in \cite{gon05} . For UHECR experiments and experimental projects, see for instance \cite{AUG,OWL,EU,NAG}. In what follows, we assume that the existence of a fundamental length scale does not alter energy and momentum conservation in the vacuum rest frame where all calculations are performed.

As we suggested in 1997, a simple LSV pattern
with an absolute local rest frame and
a fundamental length scale $a$ (e.g. the Planck scale) where new physics
is expected to occur \cite{gon97a} is given by a quadratically deformed
relativistic kinematics (QDRK) of the form \cite{gon97a, gon97b}:

\equation
E~=~~(2\pi )^{-1}~h~c~a^{-1}~e~(k~a)
\endequation
\noindent
where $h$ is the Planck constant, $c$ the speed of light, $k$ the wave vector,
and
$[e~(k~a)]^2$ is a {\bf convex}
function of $(k~a)^2$ obtained from vacuum dynamics.
Expanding equation (1) for $k~a~\ll ~1$ , we can write
in the absence of other distance and energy scales \cite{gon97b}:
\begin{eqnarray}
e~(k~a) & \simeq & [(k~a)^2~-~\alpha ~(k~a)^4~
+~(2\pi ~a)^2~h^{-2}~m^2~c^2]^{1/2}
\end{eqnarray}
\noindent
$\alpha $ being a model-dependent constant, possibly in the range $0.1~-~0.01$ for
full-strength violation of Lorentz symmetry at the fundamental length scale,
and {\it m} the mass of the particle. For momentum $p~\gg ~mc$ , we get:
\begin{eqnarray}
E & \simeq & p~c~+~m^2~c^3~(2~p)^{-1}~
-~p~c~\alpha ~(k~a)^2/2~~~~~
\end{eqnarray}
It is assumed that the Earth moves slowly with
respect to the absolute rest frame
and that calculations made in that frame apply as well, up to very small corrections, to phenomena measured from Earth or from a satellite.
The "deformation" approximated by
$\Delta ~E~=~-~p~c~\alpha ~(k~a)^2/2$ in the right-hand
side of (3) implies a Lorentz symmetry violation in the ratio $E~p^{-1}$
varying like $\Gamma ~(k)~\simeq ~\Gamma _0~k^2$ where $\Gamma _0~
~=~-~\alpha ~a^2/2$ . If $c$ is a universal parameter for all
particles, the QDRK defined by (1) - (3) preserves Lorentz symmetry
in the limit $k~\rightarrow ~0$, contrary to the standard $TH\epsilon \mu $ model \cite{will} .
QDRK can lead to several dramatic observable effects at phenomenologically reasonable energy scales \cite{gon97a,gon00a,gon00c,gon00b} , as reminded in {\bf subsections 2.1} and {\bf 2.2} . It seems to be, to date, the best-suited LSV model for phenomenology, as the value of the deformation term in the $E ~ \approx ~ 5$ x $10^{19} ~ eV$ region is reasonably close to the energy of CMB photons.

\subsection{Transition energy effects}

At energies above
$E_{trans}~
\approx ~\pi ^{-1/2}~ h^{1/2}~(2~\alpha )^{-1/4}~a^{-1/2}~m^{1/2}~c^{3/2}$,
the deformation $\Delta ~E$
dominates over the mass term $m^2~c^3~(2~p)^{-1}$ in (3) and modifies all standard kinematical balances:
physics gets closer to Planck scale than
to the electroweak scale
and ultra-high energy cosmic rays (UHECR) 
become potentially an efficient probe of Planck-scale physics. The standard parton model (in any
version) does no longer hold, and similarly for standard formulae on Lorentz
contraction and time dilation \cite{gon97d} . See, however, \cite {gon02a,gon02b} on the possible role of (formal) extra dimensions.

Because of the negative value of $\Delta ~E$ \cite{gon97e} , it costs
more and more energy, as $E$ increases,
to split the incoming longitudinal momentum in the laboratory rest frame.
As the ratio $m^2~c^3~(2~p~\Delta ~E)^{-1}$ varies like $\sim ~E^{-4}$ ,
the transition at $E_{trans}$ is very sharp. Using these simple power-like laws, QDRK can
lead \cite{gon00a, gon00b} to important observable phenomena. In particular:

- In astrophysical processes at very
high energy,
similar mechanisms can inhibit \cite{gon97d,gon00b} radiation under
external forces (e.g. synchrotron-like, where the interactions occur with
virtual photons),
photodisintegration of nuclei, momentum loss trough
collisions (e.g. with a photon wind in reverse shocks),
production of lower-energy secondaries...

- Unstable particles with at
least two stable particles in the final states
of all their decay channels become stable at very
high energy \cite{gon97a}. More generally, above $E_{trans}$, the lifetimes of all
unstable particles (e.g. the $\pi ^0$ in
cascades) become much longer than predicted
by relativistic kinematics.
The neutron or even the $\Delta ^{++}$ can be candidates for the
primaries of the highest-energy cosmic ray
events. If $c$ and $\alpha $ are not exactly universal,
many different scenarios are possible
\cite{gon97e} .

\subsection{Limit energy effects} 

$E_{trans}$ is not the only phenomenological energy scale naturally generated by DRK, and other effects are also present:

- The allowed final-state
phase space of two-body collisions is strongly
reduced at very high energy,
\cite{gon97f} ,
with a sharp fall of partial and total cross-sections
for cosmic-ray energies above
$E_{lim} ~\approx ~(2~\pi )^{-2/3}~(E_T~a^{-2}~ \alpha ^{-1}~h^2~c^2)^{1/3}$,
where $E_T$ is the target energy.
Using the
previous figures for LSV parameters, above some
energy $E_{lim} $ between 10$^{22}$ and $10^{24}$ $eV$ a cosmic
ray will not deposit most of its energy in the atmosphere
and can possibly fake an exotic event with much less energy \cite{gon97e} . Contrary to $E_{trans}$ , $E_{lim} $ depends on the target energy $E_T$ .

- For $\alpha ~a^2~>~10^{-72}~cm^2$ ,
and assuming universal values of $\alpha $ and $c$ ,
there is no GZK
 cutoff
for the particles under
consideration \cite{gon97a} . It should be noticed that, if taken in the 
$E ~ \approx ~ 5$ x $10^{19} ~ eV$ region where the GZK cutoff is expected to appear, the absence of the cutoff would also be, to some extent, a $E_{trans}$ effect as the mass term has a value close to those of the CMB photons considered. But, above this energy, it is the comparison between the deformation term and the CMB photon energy which dominates the energy balance.

- Requiring simultaneously the absence of GZK cutoff in the region
$E~\approx ~
10^{20}~eV$~, and that cosmic rays with
$E$ below $\approx ~3.10^{20}~eV$ deposit most of their energy in the
atmosphere, leads to the constraint \cite{gon97e} :
$10^{-72}~cm^2~<~\alpha ~a^2~<~
10^{-61}~cm^2$~, equivalent to $10^{-20}~<~\alpha ~<~10^{-9}$ for
$a~\approx 10^{-26}~cm$~ ($\approx~10^{21}~GeV$ scale).
Assuming full-strength
LSV forces $a$ to be in the range $10^{-36}~cm~<~a~<~10^{-30}~cm$ , but a $\approx 10^{-6}$ LSV at Planck scale
can still explain the data. Thus, the simplest
version of QDRK naturally fits
with the expected potential
role of Planck-scale dynamics if ultra-high energy cosmic rays (UHECR) are the right probe.

\subsection{QDRK with intermediate energy thresholds}

However, the simple power-like extrapolation used above for $\Delta E$ over at least 19 orders of magnitude is not the only possible behaviour of the deformation at energies below Planck scale. A simple modification would be to write:
\begin{eqnarray}
\Delta ~E~=~-~p~c~\alpha ~(k'~a)^2/2
\end{eqnarray}
\noindent
where $k' ~ = ~(k^2 + k_0^2)^{1/2} ~-~ k_0$ ~and $k_0$ is a new wavevector scale of dynamical origin associated to the energy scale $E_0 = (2\pi )^{-1}~h ~ c ~ k_0$ . For $k ~ \ll ~ k_0$ , one has: $k' ~ \approx k^2 ~ (2k_0)^{-1}$ and:
\begin{eqnarray}
\Delta ~E~=~-~p~c~\alpha ~k^4~a^2 ~k_0^{-2}/8
\end{eqnarray}
\noindent
so that the deformation becomes much smaller below the $\approx ~ k_0$ scale whereas, for $k ~ \gg ~ k_0$ , $\Delta E$ has the same form as previously up to
non-leading terms. Again, the transition between the two regimes is rather sharp. The new effective threshold scale $k_0$ ($E_0$) is to be related to some intermediate scale where new physics becomes apparent, and the parameterisation used for the new deformation is just an illustrative example. The $E_0$ scale can be chosen to be below $E_{trans}$ , between $E_{trans}$ and $E_{lim}$ or above $E_{lim}$ , leading to various phenomenological predictions.

In particular, if $k_0$ is chosen to be above the expected GZK cutoff scale, it is possible to build scenarios where the cutoff is present at the energies predicted in \cite{GZK1,GZK2} but disappears at a higher energy scale where the QDRK effects described above manifest themselves. {\bf Satellite experiments} seem to be the natural way to explore the possible existence of such a form of QDRK, irrespective of the future results of the AUGER experiment.

More generally, it would be interesting to explore scenarios where:

\begin{eqnarray}
\Delta ~E~=~-~p~c~\alpha ~(k'~a)^{\sigma }(k ~a)^{\tau }/2
\end{eqnarray}
\noindent
with $\sigma ~ + ~ \tau ~ = ~ 2$ , $\sigma $ and $\tau $ being real and positive. In this way, it is possible to further regulate the effect of the new energy scale.

\subsection{Possible energy-delayed effects}

The suppression of the GZK cutoff at the standard predicted energy can occur for a value of $\alpha $ as low as $10^{-6}$ if $a$ is the Planck length, but it may manifest itself at a higher energy if $\alpha $ is smaller. 

For $\alpha ~ \approx ~ 10^{-9}$ , the GZK cutoff would disappear at a cosmic-ray energy around 5 x $10^{20} ~ eV$ even if it is present at lower energies. A very small LSV at the Planck scale can thus be observed in the UHCR energy region. This is another important reason to carry out satellite experiments and explore cosmic-ray energies as high as possible.

~

~

\section{LDRK with a vacuum rest frame \\ and energy-momentum conservation}

LDRK, linearly deformed relativistic kinematics, was discarded in our 1997
and subsequent papers for phenomenological reasons \cite{gon00a} , but has been proposed by several authors (see e.g. \cite{amel98}) for cosmic-ray and gravitational-wave phenomenology, and various versions of
the pattern have been considered more recently \cite{Ellis2003,Amel04a}. $e~(k~a)$ is then a
function of $k~a$ and not of $(k~a)^2$
and, for $k~a~\ll ~1$ :
\begin{eqnarray}
e~(k~a) ~ \simeq ~ [(k~a)^2~-~\beta ~(k~a)^3~
+~(2\pi ~a)^2~h^{-2}~m^2~c^2]^{1/2}
\end{eqnarray}
\noindent
$\beta $ being a model-dependent constant.
For momentum $p~\gg ~mc$ :
\begin{eqnarray}
E ~ \simeq ~ p~c~+~m^2~c^3~(2~p)^{-1}~
-~p~c~\beta ~(k~a)/2~~~~~
\end{eqnarray}
\noindent
the deformation $\Delta~E~=~-~p~c~\beta ~(k~a)/2$ being now driven by an
expression linear in $k ~a$ .
LDRK can be generated by
introducing a background
gravitational field in the propagation equations of free particles \cite{el1}~.
If existing bounds on LSV from
nuclear magnetic resonance experiments are to be interpreted as setting a
bound of $\approx 10^{-21}$ on relative LSV at the momentum scale
$p~\sim ~100~MeV$ , this implies $\beta ~a~<~10^{-34}~cm$ . But LDRK seems to lead to
inconsistencies with cosmic-ray experiments unless $\beta ~a$ is much
smaller \cite{gon00a,gon00c}. Concepts and
formulae presented for QDRK can be readily extended to LDRK, and we get now:
\begin{eqnarray}
E_{trans}~
\approx ~\pi ^{-1/3}~ h^{1/3}~(2~\beta )^{-1/3}~a^{-1/3}~m^{2/3}~c^{5/3} \\
E_{lim} ~\approx ~(2~\pi )^{-1/2}~(E_T~a^{-1} \beta ^{-1}~h~c)^{1/2}~~~~~~
\end{eqnarray}
For a high-energy photon, LDRK is usually parameterized \cite{el1} as:
\begin{eqnarray}
E ~\simeq ~p~c~-~p~c~\beta ~(k~a)/2~=~p~c~-~p^2~M^{-1}
\end{eqnarray}
where $M$ is an effective mass scale. Tests of this model through
gamma-ray bursts, measuring the delays in the arrival time of photons of
different energies, have been considered in \cite{norris,Smolin05a,jac} for
the Gamma-ray Large Area Space Telescope (GLAST), and
more generally in \cite{el1} and in subsequent papers by several authors (see the references in \cite{Amel04a}).
But, from the same considerations developed in our 1997-99
papers and more systematically in \cite{gon00a,gon00c} taking QDRK as an example, stringent bounds on LDRK can be
derived.
Assume that LDRK applies only to photons, and not to charged particles, so
that at high energy we can write for a charged particle, $ch$ ,
the dispersion
relation:
\begin{eqnarray}
E_{ch} & \simeq & p_{ch}~c~+~m_{ch}^2~c^3~(2~p_{ch})^{-1}
\end{eqnarray}
\noindent
where the $ch$ subscript stands for the charged particle under consideration.
Then, it can be readily checked that the decay $ch~\rightarrow ~ch~+~\gamma$
would be allowed for $p$ above $\simeq ~(2~m_{ch}^2~M~c^3)^{1/3}$ , i.e:

- for electrons, above $E ~\approx ~2~TeV$ if $M~=10^{16}~GeV$ ,
and above $\approx ~20~TeV$ if $M~=10^{19}~GeV$ ;

- for muon and charged pions, above $E ~\approx ~80~TeV$ if $M~=10^{16}~GeV$ ,
and above $\approx ~800~TeV$ if $M~=10^{19}~GeV$ ;

- for protons above $E ~\approx ~240~TeV$ if $M~=10^{16}~GeV$ ,
and above $\approx ~2.4~PeV$ if $M~=10^{19}~GeV$ ;

- for $\tau $ leptons, above $E ~\approx ~400~TeV$ if $M~=10^{16}~GeV$ ,
and above $\approx ~4~PeV$ if $M~=10^{19}~GeV$ ;

\noindent
so that none of these particles would be observed above such energies,
apart from very short paths.
Such decays seem to be in contradiction with cosmic ray data, but
avoiding them forces the charged particles to have the same kind of
propagators as the photon, with the same effective value of $M$ up to small
differences. Similar conditions are readily derived for all
"elementary" particles, leading for all of them, up to small deviations,
to a LDRK given by the universal dispersion relation:
\begin{eqnarray}
E ~\simeq ~p~c~+~m^2~c^3~(2~p)^{-1}~-~p^2~M^{-1}
\end{eqnarray}
\noindent
For instance, $\pi^0$ production would otherwise be
inhibited. But if, as it seems compulsory, the $\pi^0$ kinematics
follows a similar law, then the decay time for $\pi^0 ~\rightarrow ~\gamma
~\gamma $ will become much longer than predicted by special relativity at
energies above  $\approx ~50~TeV$ if $M~=10^{16}~GeV$
and $\approx ~500~TeV$ if $M~=10^{19}~GeV$ . Again, this seems to
be in contradiction
with cosmic-ray data. Requiring that the $\pi^0$ lifetime agrees
with special relativity at $E ~\approx ~10^{17}~eV$ would force $M$
to be above $\approx ~10^{26}~GeV$ , far away from the values to be
tested at GLAST. Another bound is obtained from the
condition that there are $3.10^{20}~eV$ cosmic-ray events. Setting
$E_{lim}$ to this value, and taking oxygen to be the
target, yields $M~\approx ~
3.10^{21}GeV$ . It therefore
appears very difficult to make LDRK , with $M$
reasonably close to Planck scale, compatible with experimental data.

It often said that high-threshold
 ($\sim  ~ 10^{19} ~eV$) experiments like EUSO can test
"TeV gravity". By "TeV gravity" it is meant LDRK
models where the effective fundamental scale is somewhere between 10$^{16}$ and
10$^{19}$ $GeV$ . As emphasized in our previous papers \cite{gon00a,gon00b} and reminded above, present data can
already be used to exclude all versions of this LSV pattern able to lead to observable effects in the $TeV$ region (but the situation may be different for SDSR versions of LDRK). It has also been claimed that, from an experimental point of view,
the test of "TeV gravity" will be possible only after having studied ultra-high
energy neutrinos. Obviously, the study of UHE neutrinos will provide crucial
information, but there is no physical reason for such a restriction. The
condition that a UHE proton does not spontaneously decay by emitting a photon
involves only the dispersion relations of these two particles in the physical
vacuum and does not depend on any property of neutrino physics. Therefore,
"TeV gravity" based on (WDSR) LDRK patterns is ruled out by global phenomenological considerations, independently of future neutrino results.

Similarly, the discussion \cite{gon97f} of the sharp fall of multiparticle phase space
for a UHE particle interacting with the atmosphere involves only a balance between
the deformation of hadronic kinematics and the target energy which, in standard
relativistic models, is expected to provide the multiparticle transverse energy.
As a nonrelativistic target is accelerated to a relativistic speed
by the UHE collision, its rest mass turns into a much smaller mass term
($\simeq ~ m^2~p_{OT}^{-1}$ where $p_{OT}$ stands for the outgoing target momentum)
and
the released energy usually provides the transverse energy of the event as well
as the multiparticle mass terms. However, the presence of a negative deformation
term in (1) - (3) or (7), (8) growing like a power of $p$ in the kinematics 
of the incoming UHE particle alters standard
kinematical balances as studied in our
1997-2000 papers. Above $\approx ~ E_{lim}$ , there is less and less energy available to provide mass
terms and form the transverse
multiparticle phase space, so that atmospheric showers cannot be generated for
and the conventional UHECR event does no longer occur. It
is on
these and similar grounds that we excluded LDRK models long ago.

\subsection {LDRK and the bound from the Crab nebula synchrotron radiation}

Our claim that LDRK, in its standard WDSR power-like form, cannot be made consistent with experiment, has also been confirmed by the analysis astrophysical of synchrotron radiation. 
In \cite{jac03} Jacobson, Liberati and Mattingly considered the LDRK dispersion relation :
\equation
E^2(p) ~ = ~ m^2 ~ + ~ p^2 ~ + ~ \eta ~ M^{-1}
\endequation
\noindent
is considered, $M$ being the Planck mass, $\eta $ a negative constant and
$E_{QG} ~ = ~ M ~ {\vert \eta \vert }^{-1}$
the effective quantum gravity scale, to discuss data on synchrotron radiation from
the Crab nebula. They obtained a lower bound $E_{QG} ~ \gsim ~ 10^{26}GeV$. These authors
refer to \cite{gon00b} as having first pointed out the existence of a cutoff in synchrotron radiation in
the presence of Lorentz symmetry violation. In \cite{gon00b} we had made explicit the calculations for a conclusion already stated in our 1997 papers. It should also be noticed, as reminded above,
that the
bound $E_{QG} ~ \gsim ~ 10^{26} ~ GeV$ had first been obtained in \cite{gon00c} from a very reasonable requirement on $\pi^0$ lifetime .

In \cite{gon00b}, we explicitly pointed out that the transition at $E_{trans}$
introduces an essential modification
of the energy absorption required for radiation synchrotron emission and
must lead to a sharp cutoff for this emission.
Assuming $\alpha $ to be positive in (3), as
otherwise LSV would lead to spontaneous
decays at ultra-high energy \cite{gon97b,gon97e} , and using the fact that
it costs more and more energy to split the incoming longitudinal
momentum as energy increases above $E_{trans}$ , we gave a schematic illustration of the effect of DRK on synchrotron radiation using the same QDRK model as in {\bf Section 2} .
If relativistic kinematics applies, a UHE proton with energy-momentum
$\simeq ~[~p~c~+~m^2~c^3~(2p)^{-1}~ ,~p~]$ can
emit in the longitudinal direction a photon with energy
$(\epsilon ~,~ \epsilon ~c^{-1})$ if, for instance,
it absorbs an energy-momentum $\delta ~E~\simeq ~ m^2~c^2~p^{-2}~\epsilon /2$ .
This expression for $\delta ~E$ falls quadratically with the incoming energy.  
With QDRK and above $E_{trans}$ , we get instead $\delta ~E~\simeq ~
3~\alpha ~\epsilon ~(k~a)^2/2$~, quadratically rising with proton energy.
At high enough energy, the proton can no longer emit synchrotron radiation
apart from (comparatively) very small energy losses.
We therefore expect protons to be accelerated to higher energies in the
presence of Lorentz symmetry violation.
Similar considerations obviously apply to LDRK as well, where the expression

$E_{trans}~
\approx ~\pi ^{-1/3}~ h^{1/3}~(2~\beta )^{-1/3}~a^{-1/3}~m^{2/3}~c^{5/3}$ is to be compared with the cutoff $E_{max}$ obtained in \cite{jac03} which differs only by a
factor 0.93 from $E_{trans}$ and
corresponds, up to this factor, to the same analytic expression. This is
the actual origin of the cutoff rediscovered in \cite{jac03}.
A similar result for $E_{max}$ , again identical to our definition of
$E_{trans}$ , was also obtained in \cite{Ellis2003} ,
where a detailed calculation is performed using explicitly a Liouville
string model.

On the grounds of specific Liouville string models, where quantum gravity
corrections amount to introducing defects in space-time with vacuum quantum
numbers \cite{ellis00}, it was claimed \cite{Ellis2003} that charged particles may
follow a different kinematics from that of the photon, yet not spontaneously
decaying through photon emission (although kinematically allowed) as they would not see the quantum-gravitational
medium and could not emit "Cherenkov" radiation. But, although violations of the
equivalence principle by Lorentz-violating terms are obviously possible and
were already considered in our 1997 papers,
it does not appear that the claim of \cite{Ellis2003}
has actually been demonstrated and the validity of the argument is not
obvious. Quantum electrodynamics implies that a physical charged particle is
made of the bare charged particle surrounded by a cloud of virtual photons and
these photons do see the D-particle medium. The basic question is
whether a virtual photon can spontaneously materialize, and the answer seems to
be that this is kinematically allowed in the models considered in \cite{Ellis2003} . The virtual photon is indeed real if it has the required on-shell kinematical properties, and can then escape from the bare electron. The fact that the charged particle
does not see the space-time foam is not enough to contradict our assertion.
The authors of \cite{Ellis2003} compare the situation with the Cherenkov effect, but
precisely in this case the decisive ingredient is how photons see the medium
and whether their critical speed becomes lower than that of the electron.

In the application of Liouville strings considered here, particles have energies
well below Planck scale. Then, the virtual photon is comparatively long-lived and
will be emitted
in the quantum-gravitational medium even if the bare charged particle does not
see such a medium. Therefore, it does not seem that spontaneous electron decay
may be inhibited if it is kinematically allowed. Furthermore, the charged
particle can always
absorb a virtual photon from an external electromagnetic field and in this case photon emission becomes a scattering which is obvioulsy not forbidden. 

\subsection{LDRK with intermediate energy thresholds}

But, as for QDRK, it is possible to explore scenarios where $\Delta E$ is not power-like between the Planck scale and the particle mass scales. Again, we can write: $\Delta~E~=~-~p~c~\beta ~(k'~a)/2$ where $k'$ has the same definition as before. We then get for $k ~ \ll ~ k_0$:

\begin{eqnarray}
\Delta ~E~=~-~p~c~\beta ~k^2 ~ a ~k_0^{-1}/4
\end{eqnarray}
\noindent
At scales below $k_0$ , one has a form of QDRK whose exact predictions will depend on the value of $\alpha_L ~ = ~ \beta ~ (k_0 ~ a)^{-1}$ . If $k_0$ corresponds to an energy scale above the $10^{20} ~ eV$ region, then $\alpha_L ~ \approx ~ 10^{-6}$ or larger can account for the possible suppression of the GZK cutoff. If $a$ is the Planck length and $\beta ~ \approx ~ 1$ , a value of $\alpha_L$ between $10^{-6}$ and 1 implies the $E_0$ scale to lie between $\approx ~ 10^22 ~ eV$ and the Planck scale. This interval clearly includes the grand unification scale. Above $E_0$, we recover the LDRK expression (8) up to non-leading terms.

In this way, it may be possible to make QDRK phenomenology compatible with an initial LDRK pattern generated at the Planck scale.

As for QDRK, it is also possible to explore LDRK models of the form:

\begin{eqnarray}
\Delta ~E~=~-~p~c~\beta ~(k'~a)^{\sigma '}(k ~a)^{\tau '}/2
\end{eqnarray}
\noindent
with $\sigma '~ + ~ \tau '~ = ~ 1$ , $\sigma '$ and $\tau '$ being real and positive. Then, below $E_0$ , one could explore hybrid scenarios between QDRK and LDRK.
~

~

\section{Strong Doubly Special Relativity (SDSR)}

As previously quoted, other authors \cite{Amel04a,Smolin05a} require that the laws of physics be exactly the same in all inertial frames (SDSR).

We follow here the papers \cite{amel03} by Amelino Camelia et al. , and refer to the papers quoted in this article and in \cite{Amel04a,Smolin05a}. These authors
use the following deformed dispersion relation ($\lambda $ being the deformation parameter) in a two-dimensional space-time:

\begin{equation}
0= \frac{2}{\lambda^2} \left[\cosh (\lambda E)
- \cosh (\lambda m ) \right]
- p^2 e^{\lambda E}
\simeq E^2 - p^2 - m^2 - \lambda E p^2
\end{equation}
which can indeed be valid in all inertial frames at the cost of a $\lambda$-dependent
deformation of the boost transformations (SDSR), but can also be interpreted as being valid only in a preferred reference system (WDSR).
Amelino-Camelia et al. emphasize that this formulation of SDSR is not compatible with standard energy-momentum conservation.
To show this incompatibility, they use
the dependence of energy-momentum on
the rapidity parameter $\xi$ :
\begin{equation}\label{xidsr1}
\cosh (\xi) = \frac{e^{\lambda E} - \cosh\left(\lambda m\right)}
  {\sinh\left(\lambda m \right)} ~,~~~
\sinh (\xi) = \frac{\lambda p e^{\lambda E}}
  {\sinh\left(\lambda m\right)}\,\, ,
\end{equation}
where $\xi$ here is the amount of rapidity needed to take a particle
from its rest frame ($E=m$, $p=0$)
to a frame in which its energy is $E$ and its momentum is 
$p(E)$ from the dispersion relation (17).
For $\lambda \rightarrow 0$ , one gets
the standard special relativistic relations:
\begin{equation}
\cosh (\xi) = \frac{E}{m} ~,~~~
\sinh (\xi) = \frac{p}{m} \,\, .
\end{equation}
Amelino-Camelia et al. check that, if one is to enforce the standard additive law of energy-momentum conservation
in a framework where (17) and (18) hold in all inertial frames,
such a law can
only hold in one inertial frame. They conclude that this cannot
be the SDSR law of energy-momentum conservation and, referring to previous authors, propose a conservation law whose form to first order in $\lambda $ is:
\begin{equation}
E_a + E_b - \lambda p_a p_b
- E_c -E_d + \lambda p_c p_d = 0
~,
\end{equation}
\begin{equation}
p_a + p_b - \lambda (E_a p_b + E_b p_a)
- p_c -p_d + \lambda (E_c p_d + E_d p_c) = 0
\end{equation}
\noindent
Such a conservation law means in particular that {\bf energy and momentum are not
additive}. Therefore, we are not dealing with free particles strictly speaking, as the deformation generates an effective interaction between the "free particle" and the other particles in the Universe. Refering to \cite{Hey} , Smolin \cite{Smolin05a} describes this situation by saying that {\it "while it is the covariant energy and momentum which are observed, it is the contravariant 4-vectors which are conserved linearly"}.

The fact that the particles under
consideration are not really free raises the question of whether they all see
the same space-time properties as they propagate in a SDSR frame. Furthermore, how the define the particle velocity? It has been recognized \cite{Kow04} that there is a "spectator problem" in such an approach, as any particle in the Universe interacts with all the other particles and the hamiltonian involves kinematical terms relating each single particle to the rest of existing matter. But then, velocity should be defined in terms of the global hamiltonian and not of the formal hamiltonian of an isolated particle which has no real physical meaning. To date, there seems to be no clear solution to this possible source of fundamental inconsistency.

Actually, to consistently define the velocity of a single free particle, one must be able to separate its individual hamiltonian from that of the rest of the world. This does not seem to be possible in SDSR. By willing to preserve the strict universality of the laws of Nature in all inertial frames, an old fundamental principle is abandoned: that of the separability of a free particle from the rest of matter. It thus seems impossible to determine the velocity of a single "free" particle without knowing the existence and motion of all matter in the Universe. The matter motion and distribution observed will in principle depend on the inertial frame, so that SDSR may naturally generate its own breaking in the real universe. Perhaps this is a strong indication that, in the real world, no consistent departure from the standard Lorentz symmetry can afford itself working without a preferred inertial frame.

It is well known that there is in SDSR another possible conservation law, obtained defining the physical energy and momentum $E'$
and $p'$ such that:
\begin{equation}
\frac{E'}{m} = \frac{e^{\lambda E} - \cosh\left(\lambda m\right)}
  {\sinh\left(\lambda m \right)} ~,~~~
\frac{p'}{m}  = \frac{\lambda p e^{\lambda E}}
  {\sinh\left(\lambda m\right)}\,\, ,
\end{equation}
so that $E'$ and $p'$ are additive and conserved. With these two variables and
standard definitions of space and time, we readily recover special relativity. This may raise a conceptual puzzle for SDSR time of travel tests. If each inertial SDSR reference frame can be related to a (non physical) standard inertial frame of special relativity (the SR frame), two photons of different energies emitted simultaneously and at the same position in the SR frame will follow identical paths in a relevant system of local SR frames and arrive simultaneously to the detector in its local SR rest frame. Then, a macroscopic difference in arrival time cannot be generated by the local transformation between the SR and the SDSR frame at the detector position.

The requirement that the laws of physics be exactly the same in all inertial frames naturally implies that most usually suggested tests of LSV are not suitable for the "strict" interpretation of DSR advocated by Amelino-Camelia et al. \cite{amel03} and other authors. According to their study, the possible tests of SDSR through observable effects seem to be essentially those based on time-of-travel experiments (but there may be a nontrivial problem in determining the speed of a "free" particle). This result is to be compared with our previous analysis \cite{gon02a,gon02b}, where we pointed out that actually the Kirzhnits-Chechin ansatz (KCh) was not able to
reproduce the GZK cutoff. As previously stressed, the KCh model is nothing but a form of SDSR, where it is required that the Finsler space law replaces special
relativity in all inertial frames. Thus, more recent references generalize our 2002 result and confirm that our 1997 papers proposed the best suited LSV pattern to explain possible unconventional results in UHECR experiments.

\subsection{SDSR with intermediate energy thresholds}

No basic principle prevents patterns of the SDSR type from presenting the same kind of intermediate energy thresholds considered in subsections {\bf 2.1} and {\bf 3.1} . Then, the absence of a measurable effect in time of travel experiments would not necessarily provide a clear way to falsify SDSR. But if the effect is found and turns out to have no "natural" explanation, it may provide a serious evidence for SDSR, as the WDSR pattern would lead to many unwanted phenomena in this case.
~

~

\section{On models involving superluminal particles}

The idea that conventional Lorentz symmetry could  be only an approximate
property of equations describing a sector of matter, and that it would  
be broken at very high energy and short distance, was already put forward
in our 1995-96 papers on superluminal particles \cite{gon95}. This kind of
Lorentz symmetry breaking, due to the mixing with superluminal sectors,
had then to be a general property of the equations of the "ordinary" sector
of matter, including propagators and dispersion relations deformed by the Lorentz breaking mixing. 

In \cite{gon95} it
was also pointed out that superluminal particles with positive mass and
energy (superbradyons) must necessarily emit "Cherenkov" radiation, i.e.
particles with lower critical speed in vacuum. This was the basic property
used later in \cite{glash} to claim bounds on models of the $TH \epsilon \mu $ type with non-universal critical speed in vacuum.
Similarly, in \cite{gon95} we already suggested scenarios of Lorentz symmetry
Violation, with superluminal particles, allowing to escape the GZK cutoff.

In \cite{gonsl97}, we also considered for the first time: a) models breaking
simultaneously Lorentz symmetry and the symmetry between particles and
antiparticles, as well as the possibility that this mechanism explains
the difference between matter and antimatter in the Universe; b) specific DRK patterns generated by the mixing with superluminal particles. More recent papers are \cite{gonsl03} .

As an example, the scale corresponding to the rest energy of a superluminal particle (or a family of them) can naturally set an intermediate energy scale for the energy dependence of the deformation term in DRK patterns. More involved mechanisms can also be considered.

~

~
\section {Conclusion}

{\bf The AUGER experiment alone cannot completely settle the most crucial issues of UHECR phenomenology, and must be completed by UHECR space experiments}. These experiments should 
not only be sensitive to the highest possible cosmic-ray energies and to the lowest possible fluxes, 
but must have at the same time an energy threshold as low as possible. This second requirement reflects the need to understand the first cosmic-ray interactions in the atmosphere, as well as the beginning of cascade development. The $\pi ^0$ lifetime at UHE, for instance, is a crucial parameter, but it is far from being the only one. The validity of standard dynamics at energies which are closer to the Planck scale than to accelerator scales is by itself a fundamental issue, especially in the presence of LSV.

The question of whether a vacuum rest frame exists in our Universe remains to be answered by experiment. The absence of the GZK cutoff would be a clear indication against SDSR and in favour of WDRS with a vacuum rest frame, unless a more conventional explanation could be found.

If the GZK cutoff turns out to exist below $10^{20} ~eV$ , this will not rule out all possible WDSR patterns. Models with an intermediate energy scale in the UHECR region, as well as delayed-energy scenarios, will have to be explored and checked. Thus, satellite experiments are in any case necessary independently of the future AUGER results.

~

~

\end{document}